\documentclass[twocolumn,showpacs,preprintnumbers,amsmath,amssymb]{revtex4}

\usepackage{graphicx}
\usepackage{dcolumn}
\usepackage{bm}
\usepackage{upgreek}

\begin{document}


\title{High-fidelity quantum operations on superconducting qubits in the presence of noise}

\author{Andrew J. Kerman and William D. Oliver}
\affiliation{Lincoln Laboratory, Massachusetts Institute of
Technology, Lexington, MA, 02420}

\date{\today}

\begin{abstract}

We present a scheme for implementing quantum operations with
superconducting qubits. Our approach uses a ``coupler'' qubit to
mediate a controllable, secular interaction between ``data'' qubits,
pulse sequences which strongly mitigate the effects of 1/f flux
noise, and a high-Q resonator-based local memory. We develop a
Monte-Carlo simulation technique capable of describing arbitrary
noise-induced dephasing and decay, and demonstrate in this system a
set of universal gate operations with $\mathcal{O}(10^{-5})$ error
probabilities in the presence of experimentally measured levels of
1/f noise. We then add relaxation and quantify the decay times
required to maintain this error level.
\end{abstract}

\maketitle

Superconducting qubits are promising building blocks for quantum
computers~\cite{Makhlin01a}. Realizing this promise requires a qubit
coupling scheme and associated control pulse sequences capable of
realizing gate operations with error probabilities low enough to
achieve fault-tolerance~\cite{thresh}. Such a control scheme should
ideally permit a practical, time-efficient implementation of
universal single- and two-qubit operations; incorporate switchable
coupling between qubits; and be insensitive both to decoherence and
fabrication variations of the qubits.

Although several
proposed~\cite{doubleres,Niskanen06a,para,couplingtheory} and
experimental~\cite{couplingexp} qubit coupling schemes already
exist, none yet combine all of the above features. In this Letter,
we present a scheme which aims to do this. Our approach utilizes a
``coupler'' qubit to mediate interaction between ``data"
qubits~\cite{Niskanen06a}, pulse sequences incorporating spin-echoes
to suppress the effects of 1/f flux
noise~\cite{Yoshihara06a,Kakuyanagi07a}, and local memory in the
form of high-Q resonators~\cite{Koch06a}. We introduce a Monte-Carlo
technique capable of simulating gate operations in the presence of
noise and spontaneous decay, which we then use to show that
$\mathcal{O}(10^{-5})$ error probabilities (comparable to certain
predicted thresholds for fault-tolerance~\cite{thresh}) can be
achieved in the presence of experimentally measured levels of flux
noise. Finally, we quantify the qubit excited-state lifetimes and
control signal requirements necessary to achieve this.

\begin{figure}
\includegraphics[width=3.25in]{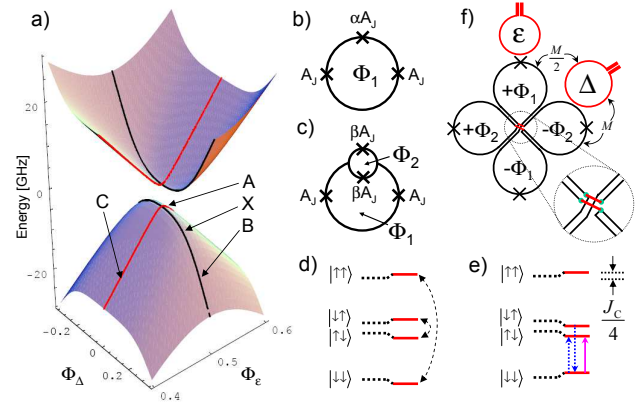}
\caption{\label{fig:1} (color online) Flux qubits and secular vs.
nonsecular coupling. (a) Lowest-two energy levels for a flux qubit.
Schematic of (b) single-loop 3-JJ and (c) double-loop 4-JJ flux
qubits. Relative junction areas have ratios $\alpha$ and
$\beta$~\cite{Orlando99a}. Single-loop qubits have only one control
flux $\Phi_\epsilon$ and are restricted to the line AC in (a).
Double-loop qubits have two control fluxes:
$\Phi_\epsilon\equiv\Phi_1+\frac{\Phi_2}{2}$ and
$\Phi_\Delta\equiv\Phi_2$. (d) Energy levels for two flux qubits at
degeneracy, with nonsecular coupling. Level shifts result from
second-order mixing of two-qubit eigenstates (dashed arrows). (e)
Energy levels for secular coupling. A conditional level shift allows
CNOT to be implemented with a single microwave transition (solid
arrow), and SWAP with a stimulated Raman transition (by driving a
2$\pi$ pulse on each qubit, corresponding to a $\pi$ pulse on the
Raman transition). (f) possible layout of a gradiometric 4-JJ qubit,
for which $\Phi_1$ and $\Phi_2$ can be accessed independently. The
$\Delta$ control-loop orientation induces $\Phi_1$ and $\Phi_2$ in
the -0.5:1 ratio required to adjust $\Delta$ without affecting
$\epsilon$.}
\end{figure}

We focus here on the flux qubit~\cite{Orlando99a}, although much of
what follows is more generally applicable. In most work to date,
this consists of a single superconducting loop interrupted by
several Josephson junctions [Fig. \ref{fig:1}(b)]. In this work,
however, we consider the double-loop four-junction version [Fig.
\ref{fig:1}(c)]. When $\Phi_1,\Phi_2 \approx \frac{\Phi_0}{2},0$
($\Phi_0=h/2e$ is the flux quantum), the qubit can be described by
the approximate two-level Hamiltonian
$\hat{H}_\textrm{qb}=\frac{1}{2}[\epsilon\hat\sigma^z+\Delta\hat\sigma^x]$.
The eigenstates of $\hat{\sigma}^\textrm{z}$ are ``persistent
current" (PC) states with energies $\pm\epsilon/2 = \pm
I_{\textrm{p}} (\Phi_1 - \frac{\Phi_0}{2})$, where currents $\pm
I_{\textrm{p}}$ circulate around the $\Phi_1$ loop. The parameters
$\epsilon$ and $\Delta$ are tunable by the fluxes
$\Phi_{\epsilon}\equiv\Phi_1+\frac{\Phi_2}{2}$ and $\Phi_{\Delta}
\equiv \Phi_2$ [Figs.~\ref{fig:1}(c),(f)]. Point A in
Fig.~\ref{fig:1}(a), where the PC states are mixed to produce an
avoided crossing with $\hat{\sigma}^x$-like energy eigenstates, is
known as the ``degeneracy point''. It has the desirable feature that
$dE / d\Phi_{\epsilon}=dE / d\Phi_{\Delta}=0$ (where the qubit
energy splitting $E=\sqrt{\epsilon^2 + \Delta^2}$) so that
decoherence due to flux noise is minimized~\cite{Yoshihara06a}.
Existing qubit coupling schemes which are insensitive to decoherence
rely on biasing at this point~\cite{doubleres,para,Niskanen06a}.

There are, however, two disadvantages to working at degeneracy:
first, $E$ is then fixed by fabrication and may vary significantly
between qubits, requiring individually tuned microwaves; second, the
degenerate qubit is first-order insensitive to flux from other
qubits, so that inductively coupling $\Phi_1$-loops, as in previous
works~\cite{doubleres,para,Niskanen06a}, has no effect on the system
to first order. This ``nonsecular" coupling can be written in the
form $\hat{H}_\textrm{C}^\textrm{zz}\equiv
J_{\textrm{C}}^\textrm{zz}\hat\sigma^\textrm{z}_1\hat\sigma^\textrm{z}_2$
(the subscripts indicate qubits 1 and 2), and its leading-order
effect at degeneracy is a second-order mixing [Fig. \ref{fig:1}(d)].
This mixing can be exploited using double resonance \cite{doubleres}
or parametric driving \cite{para,Niskanen06a} to enable, for
example, the two-qubit $\sqrt{\textrm{SWAP}}$ gate, even if the
qubits have different splittings \cite{doubleres,para,Niskanen06a};
however, these schemes require different microwave frequencies for
each qubit or qubit pair. Furthermore, the Controlled-NOT (CNOT)
gate used ubiquitously in quantum circuits must then be constructed
from several $\sqrt{\textrm{SWAP}}$ and single-qubit
gates~\cite{SWAP}.


\begin{figure}
\includegraphics[width=3.25in]{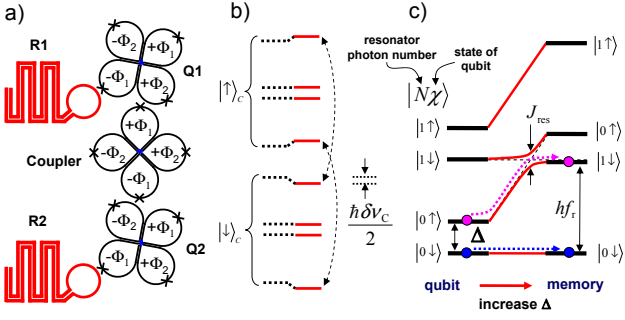}
\caption{\label{fig:2} (color online) Possible scheme for universal
two-qubit operations. (a) A 4-JJ coupler qubit mediates the
interaction between two 4-JJ data qubits (Q1, Q2). Coupling is
switched off by tuning $\epsilon_\textrm{C}\gg\Delta_\textrm{C}$.
(b) An inductive interaction mixes the coupler levels (dashed
arrows), resulting in a conditional frequency shift
$\delta\nu_\textrm{C}$ in the computational subspace $\mathcal{C}_2$
[see Fig. \ref{fig:1}(e)]~\cite{parity}. (c) Increasing the qubit
$\Delta$ adiabatically through the avoided crossing transfers the
qubit state to the resonator~\cite{Koch06a}.}
\end{figure}

Far from degeneracy ($\epsilon\gg\Delta$), however, the energy
eigenstates are approximately PC states, and
$\hat{H}_\textrm{C}^\textrm{zz}$ commutes with
$\hat{H}_\textrm{qb}$, giving a first-order level shift [Fig.
\ref{fig:1}(e)] which lifts the degeneracy between the
$|\downarrow\downarrow\rangle\leftrightarrow|\uparrow\downarrow\rangle$
and
$|\downarrow\uparrow\rangle\leftrightarrow|\uparrow\uparrow\rangle$
transitions by the conditional frequency shift
$\hbar\delta\nu_\textrm{C}=J_{\textrm{C}}^\textrm{zz}$. This
``secular" coupling allows CNOT to be driven with a single microwave
pulse (Fig. \ref{fig:1}(e), solid arrow), and SWAP with a stimulated
Raman transition (dashed arrows). However, the qubits are now
sensitive to flux noise due to the large slope
$dE/d\Phi_\epsilon=2I_\textrm{p}$.

Based on these observations, we consider the system shown in
Fig.~\ref{fig:2}(a) consisting of two two-loop ``data'' qubits,
whose coupling is mediated by a third ``coupler'' qubit in a manner
similar to that used in Ref.~\onlinecite{Niskanen06a} for
single-loop qubits. Here, however, we inductively couple to
$\Phi_\Delta$ of each data qubit rather than $\Phi_1$ (as in
previous works), with the Hamiltonian:
$\hat{H}_\textrm{C}=J_\textrm{C}\hat\sigma_\textrm{C}^\textrm{z}
(\hat\sigma_1^\textrm{x}+\hat\sigma_2^\textrm{x})$, where
$J_\textrm{C}\approx M_\textrm{iC}I_\textrm{p}
d\Delta_\textrm{i}/d\Phi_S$ and $M_\textrm{iC}$ is the mutual
inductance between data and coupler qubits \cite{edeltaratio}. When
$\epsilon_1=\epsilon_2=\epsilon_\textrm{C}=0$, this configuration
produces the level structure shown in Fig.~\ref{fig:2}(b)
\cite{chargealso}. The lower four of these eight eigenstates
(hereafter the two-qubit ``computational subspace" or
$\mathcal{C}_2$) exhibit an effective two-qubit conditional
frequency shift
$\hbar\delta\nu_\textrm{C}=\Delta_C\{1-\sqrt{1+(2J_\textrm{C}/\Delta_\textrm{C})^2}\}$
equivalent to that shown in Fig.~\ref{fig:1}(e)~\cite{parity}. In
addition, this shift can be turned off by tuning
$\epsilon_\textrm{C}/\Delta_\textrm{C}\gg1$, so that the coupler
eigenstates become approximately PC states (and eigenstates of
$\hat{H}_\textrm{C}$); this suppresses the mixing shown in Fig.
\ref{fig:2}(b), and with it $\delta\nu_\textrm{C}$ by the factor
$\sim(\Delta_\textrm{C}/\epsilon_\textrm{C})^3$. The speed of this
switching is limited only by nonadiabatic excitation of the coupler
qubit~\cite{appcheck}. Finally, the sensitivity of each data qubit
to flux noise in this configuration is given by
$d\Delta/d\Phi_\Delta=J_\textrm{C}/MI_\textrm{p}^\textrm{C}$
($I_\textrm{p}^\textrm{C}$ is the coupler persistent current), which
can be reduced substantially by using larger junctions in the
coupler qubit to increase $I_\textrm{p}^\textrm{C}$.

The final ingredient shown in Fig.~\ref{fig:2}(a) are the
transmission-line resonators coupled to the $\Phi_\epsilon$-loop of
each qubit so that its state can be adiabatically transferred into a
resonator state \cite{Koch06a} [Fig.~\ref{fig:2}(c)]. These
resonators provide a potentially long-lived memory in which qubit
states can be stored between manipulations.

To simulate our system, we integrate the time-dependent
Schr\"{o}dinger equation with the Hamiltonian:

\begin{align}
\hat{H} = &\frac{1}{2}\sum \left \{ \left
[\epsilon_i(t)+\hbar\Omega_{\upmu
    i}(t)\cos(\omega_it)\right ]\hat\sigma_i^\textrm{z}
    +\Delta_i(t)\hat\sigma_i^\textrm{x} \right \}\nonumber \\
     & +\sum \left
    [hf_{\textrm{r}j}\hat{a}_{\textrm{r}j}^\dagger\hat{a}_{\textrm{r}j}+J_\textrm{res}
    (\hat{a}_{\textrm{r}j}^\dagger+\hat{a}_{\textrm{r}j}^{})\hat\sigma_j^\textrm{z}\right
    ] +\hat{H}_\textrm{C}\label{eq:Htot}
\end{align}
\noindent where $i\in\{1,2,\textrm{C}\}$, $j\in\{1,2\}$, and we have
added microwave fields illuminating the $\Phi_\epsilon$ loop of each
qubit. We treat the qubits as two-level systems and truncate the
ladder of resonator photon states at $n=1$~\cite{appcheck}. As fixed
input parameters, we assume resonator frequencies $f_{r1,r2}=9,11$
GHz (they may be identical in general), qubit-resonator coupling
$J_\textrm{res}/h= 0.75$ GHz, and qubit-qubit coupling
$J_\textrm{C}/h= 0.3$ GHz~\cite{numvals}.

\begin{figure*}
\includegraphics[width=7in]{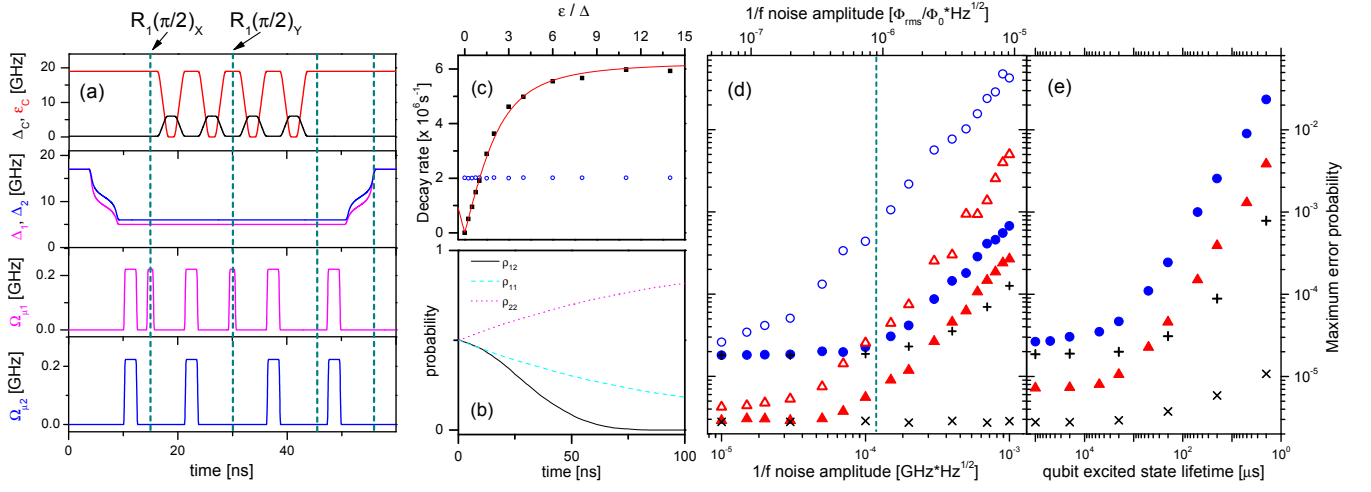}
\caption{\label{fig:3} (color online) Simulation of gate operations
in the presence of noise. (a) CNOT pulse sequence. Data qubits are
ramped to $\Delta_{1,2}=$5,6 GHz during operations, e.g., point X
[Fig.~\ref{fig:1}(a)] ($\Delta_{1,2}$ are chosen to be different
here only for clarity); the coupler has
$(\epsilon_\textrm{C},\Delta_\textrm{C})=(17,0.2)$ GHz and $(0,4)$
GHz in the decoupled and coupled states,
respectively~\cite{numvals}. Pulsed microwaves generate $\pi$ and
$\pi/2$ rotations; $C\pi/4$ operations are generated by coupling for
a controlled duration. Spin-echoes are generated at times indicated
by dashed lines using $\pi$-pulses. (b) Simulation of a single qubit
including 1/f noise and spontaneous decay (see text). (c) Decay
rates as a function of $\epsilon/\Delta$. Solid symbols show the
fitted exponential decay rate $\Gamma_\textrm{e}$ of $\rho_{22}$
(constant by construction). Open symbols show the decay rate
$\Gamma_\textrm{g}$, extracted by fitting $|\rho_{12}|$ to the
approximate expression: exp$\left [
-\Gamma_\textrm{e}t-(\Gamma_\textrm{g}t)^2\right ]$ with
$\Gamma_\textrm{e}$ fixed at the fitted value. The solid red line is
based on the theory in Ref. \cite{ithier05} with no free parameters.
(d) maximum gate error $\mathcal{E}$ for H (triangles) and CNOT
(circles) gates with $T_1\rightarrow\infty$ and noise spectral
density $S(\omega)=A_\omega/\omega$, vs. $A_\omega$. Open symbols:
same operations without spin-echo compensation. Vertical dashed
line: experimentally measured noise level of 1
$\mu\Phi_0/\sqrt{\textrm{Hz}}$ at 1 Hz~\cite{Yoshihara06a}. (e)
Maximum error probabilities at that noise level versus the qubits'
excited state lifetime. In both (d) and (e), (+) and ($\times$) show
$\mathcal{E}$ for CNOT and H, respectively, with noise and decay
\textit{on the coupler qubit only}.}
\end{figure*}

We have simulated Hadamard (H), CNOT, and SWAP gates. As an example,
the CNOT pulse sequence is shown in Fig. \ref{fig:3}(a). Both qubits
begin with states stored in their respective resonators
($\Delta_{1,2}$ high), and with the coupler ``off"
($\epsilon_\textrm{C}/\Delta_\textrm{C}\gg 1$). A downward ramp is
applied to both $\Delta_i$ to retrieve the qubits' states (only one
$\Delta_i$ would be ramped for H). These ramps are optimized to
maintain a constant non-adiabaticity as the avoided crossing in Fig.
\ref{fig:2}(c) is traversed. Next, the qubits are manipulated with
single-qubit microwave rotations and conditional phase (C$\Phi$)
gates produced by transiently pulsing the coupler ``on"
$(\epsilon_\textrm{C}/\Delta_\textrm{C}=0)$ using linear flux
ramps~\cite{appcheck}. The CNOT pulse sequence (with bit 1 the
target qubit) shown in Fig.~\ref{fig:3}(a) is:
$R_{1,2}(\pi)R_1^x(\frac{\pi}{2})C^e(\frac{\pi}{2})R_1^y(\frac{\pi}{2})C^e(\frac{\pi}{2})R_{1,2}(\pi)$,
where $R_{p,q}^\nu(\theta)$ is a rotation of angle $\theta$ around
axis $\nu$ on the Bloch sphere for qubits $p$ and $q$,
$C^e(\frac{\pi}{2})\equiv
C(\frac{\pi}{4})R_{1,2}(\pi)C(\frac{\pi}{4})$, and we have
suppressed the free evolution intervals for clarity. The
$\pi$-pulses induce spin-echoes at the center of each $\pi/2$-pulse
and at the end of the operation. H and SWAP can be constructed
similarly: $H=R_{1}(\pi)R_i^x(\frac{\pi}{2})R_{1}(\pi)$, and
$\textrm{SWAP}=
R_{1,2}(\pi)R_{1,2}^x(\frac{\pi}{2})C^e(\frac{\pi}{2})R_{1,2}^y(\frac{\pi}{2})
C^e(\frac{\pi}{2})R_{1,2}^x(\frac{\pi}{2})C^e(\frac{\pi}{2})R_{1,2}(\pi)$.
Finally, the qubit states are transferred back into the resonators,
and the final relative phase of each qubit is adjusted by varying
the time at which this occurs.

All microwave pulses have a fixed amplitude (with 100-ps rise/fall
times), so that pulse area is controlled solely by their duration.
This is crucial for producing effective spin-echoes with short,
intense pulses at a fixed frequency, due to the Bloch-Siegert shift
\cite{bsshift}. This frequency shift is amplitude dependent, so the
microwave amplitude must be kept fixed in order to compensate for it
with a fixed frequency offset (here, $\sim$5 MHz).

The use of single-qubit rotations and C$\Phi$ gates instead of the
frequency-resolved pulses shown in Fig.~\ref{fig:1}(e) has several
advantages: first, it requires only a single microwave frequency for
all operations (assuming all qubits are tuned to the same frequency
using $\Phi_\Delta$); second, it does not require accurately
resolving two frequencies spaced by $\delta\nu_\textrm{C}$ using
long microwave pulses. Finally, all gate parameters can be adjusted
solely by varying the \textit{timing} of the microwave and flux
pulse edges. These characteristics may also be of interest for other
types of qubits.

We simulate the effect of decoherence due to flux noise using a
Monte-Carlo technique. Noise is added to each $\Delta_i$ and
$\epsilon_i$ in eq.~\ref{eq:Htot} by superposing an independent,
discrete set of random two-level ``telegraph''
signals~\cite{frequencies}, which produce a 1/f average spectral
density. For each Monte-carlo iteration (all of which can be run in
parallel), a new set of noise functions is generated (one for each
$\epsilon,\Delta$). From these we obtain a sequence of $N$ time
intervals $\{\delta\tau_\textrm{i}\}$ (typically $N\sim$ several
hundred) whose union is the full gate time, and over each of which
all fluctuators are \textit{constant}. We then integrate the 32
coupled Schr\"{o}dinger equations sequentially over these intervals.
The final states for all iterations are converted to density
matrices and then averaged, corresponding to a statistical mixture
of the different noise realizations.

Spontaneous decay is added as follows. At the end of each
$\delta\tau_i\equiv t_i-t_{i-1}$, we construct the density matrix
$\rho_i$ and apply amplitude damping to the qubits:
$\rho_i^\gamma=\mathcal{S}^1(\gamma_i)\mathcal{S}^2(\gamma_i)\mathcal{S}^\textrm{C}(\gamma_i)\rho_i$
where $\mathcal{S}^j(\gamma)\rho=\sum_k U_k^j(\gamma)\rho
U_k^j(\gamma)^\dagger$, and:
\begin{equation}
U_{1,2}^j(\gamma)=M_\textrm{d}^j\left\{ \left(
\begin{array}{ccc}
1 & 0 \\
0 & \sqrt{1-\gamma} \end{array} \right),\left(
\begin{array}{ccc}
0 & 0 \\
\sqrt{\gamma} & 0 \end{array}
\right)\right\}(M_\textrm{d}^j)^\dagger.
\end{equation}
\noindent Here, $\gamma_i=1-\textrm{exp}[-\delta\tau_i/T_1^j]$ is
the probability of decay for qubit $j$ during $\delta\tau_i$,
$T_1^j$ is the excited-state decay time for qubit $j$, and
$M_\textrm{d}^j$ transforms from the instantaneous energy eigenbasis
of the \textit{isolated} qubit $j$ at time $t_i$ (decay is assumed
to occur in this basis) to the fixed basis of the
simulation~\cite{coop}. To return to state-vector space at the end
of each $\delta\tau_i$, we first write
$\rho_i=\sum_k\lambda_i^k|\psi_i^k\rangle\langle\psi_i^k|$. In
principle, we could then pick one normalized $|\psi_i^k\rangle$ at
random according to the probabilities $\lambda_i^k$; however, since
$\lambda_i^k\leq\mathcal{O}(\gamma_i)$ for $k\neq 1$ (with
$\lambda_i^k$ in decreasing order for given $i$), convergence of the
Monte-Carlo sum would require many iterations to sample any of these
unlikely events. Instead, we use an expansion procedure. For each
noise realization, we run two simulations; for the first, we choose
$|\psi^1_i\rangle$ at every time $t_i\in \{t_1\cdots t_N\}$. This
corresponds to the leading term in a weighted average, having
probability
$\mathcal{P}_0\equiv\prod_i\lambda_i^1\sim\mathcal{O}(1-t_N/T_1)$.
Next, we use the $\{\lambda_i^k\}$ obtained from simulation of the
first term to obtain approximate relative probabilities
$\mathcal{P}(m,n)$ for the next order terms, sequences of
eigenvectors of the form
$\{\psi_1^1,\cdots,\psi^1_{n-1},\psi^m_n,\psi^1_{n+1},\cdots,\psi^1_N\}$
with  $m\neq 1$ (exactly one unlikely event), as:
$\mathcal{P}(m,n)\approx\lambda_1^1\cdots\lambda_{n-1}^1\lambda_n^m$
(we neglect the possibility of two or more unlikely events); we then
randomly pick one of these terms according to normalized probability
distribution: $\mathcal{P}(m,n)/\sum_{m,n}\mathcal{P}(m,n)$, and
simulate it. Finally, we combine the density matrices for these two
simulations with the relative weights $\mathcal{P}_0$ and
$\sum_{m,n}\mathcal{P}(m,n)\approx 1-\mathcal{P}_0$.

Figures ~\ref{fig:3}(b) and (c) show a test of this method on a
single qubit initially in the state
$(|\uparrow\rangle+|\downarrow\rangle)/\sqrt{2}$, under the
influence of both 1/f noise and decay. As expected, the excited
state decays exponentially, while the coherence does not
[Fig.~\ref{fig:3}(b)]. Decay rates can be extracted from these data,
and are shown in Fig. \ref{fig:3}(c) as a function of
$\epsilon/\Delta$. The solid line is a prediction, with no free
parameters, using the theory from Ref.~\onlinecite{ithier05}.

To characterize a gate operation, we run the simulation (in
parallel) for each of a set of chosen initial states
$|\phi_0^\nu\rangle$ spanning $\mathcal{C}_2$. From these outputs we
can reconstruct the superoperator $\mathcal{S}_\textrm{op}$ for the
gate (acting on density matrices in $\mathcal{C}_2$), which can be
represented as a 16 x 16 matrix acting on ``vectors'' obtained by
stacking the columns of the 4 x 4 density matrix~\cite{QPT}. For
decoherence only (no decay), each iteration is a linear operation,
and only four $|\phi_0^\nu\rangle$ are required. With decay,
however, this is no longer true due to the projection and
renormalization, and $\textit{sixteen}$ $|\phi_0^\nu\rangle$ are
required.

We calculate the maximum error probability: $\mathcal{E}=1-
\textrm{min}\left
[\langle\Psi|\mathcal{S}_0^{-1}\mathcal{S}_\textrm{op}\rho|\Psi\rangle\right
]$, where $\mathcal{S}_0$ is the superoperator corresponding to the
desired operation, and maximization is over normalized state vectors
$|\Psi\rangle$ in $\mathcal{C}_2$. Figure~\ref{fig:3}(d) shows
$\mathcal{E}$ for the CNOT (circles) and Hadamard (triangles) gates,
both with spin-echo compensation [fig. \ref{fig:3}(a)] (filled
symbols) and without it (open symbols), as a function of 1/f-noise
amplitude $A_\omega$, for
$T_1\rightarrow\infty$~\cite{resolution,nonad}. The spin-echo
compensation strongly suppresses errors due to 1/f noise, keeping
$\mathcal{E}$ at or below $\mathcal{O}(10^{-5})$ up to the noise
level observed in Ref.~\onlinecite{Yoshihara06a}, indicated by the
vertical dashed line. Figure~\ref{fig:3}(e) shows $\mathcal{E}$ for
this same noise level as a function of the qubits' excited state
lifetime, which would evidently have to be $\sim$ ms for
$\mathcal{E}\sim 10^{-5}$. Finally, (+) and ($\times$) show
$\mathcal{E}$ with noise and decay added only to the coupler qubit,
illustrating that although the extra qubit does in principle expose
the system to more noise, the present configuration is insensitive
to it.

Although $\mathcal{E}$ can be used to set bounds on fault-tolerance
thresholds for quantum circuits, the $\mathcal{S}_\textrm{op}$
generated by these simulations give much more: a full statistical
description of the errors that occur for an arbitrary input state.
This may provide a means to compute more accurate thresholds for
real quantum circuits (assuming that noise correlations between
successive operations can be neglected). Furthermore, it may enable
the design of optimized error-correction schemes and/or gate
operation protocols which target a particular noise source, enabling
higher thresholds for fault-tolerance.

This work is sponsored by the United States Air Force under Air
Force Contract \#FA8721-05-C-0002.  Opinions, interpretations,
recommendations and conclusions are those of the authors and are not
necessarily endorsed by the United States Government.


\begin{thebibliography}{}
\bibitem{edeltaratio} Note that a deviation $\delta_M^i$ from the -0.5:1 ratio in Fig. \ref{fig:1}(f)
gives a term
$\delta\hat{H}_\textrm{C}=\delta_M^iJ_\textrm{C}\hat\sigma_\textrm{C}^\textrm{z}
\hat\sigma_i^\textrm{z}$, whose leading effect arises in the
qubit-resonator state transfer. Our simulations show that
$\delta_M^i<0.5\%$ maintains errors of $\mathcal{O}(10^{-5})$.
Slowing down the transfer or removing the memory entirely leaves the
next order effect from the $C\Phi$ gates, where $\sim$ten times
larger $\delta_M^i$ gives $\mathcal{O}(10^{-5})$ errors.
\bibitem{chargealso} The same structure could also be obtained
by coupling split cooper-pair box qubits with a flux qubit.
\bibitem{parity} The level structure in Fig.~\ref{fig:2}(b) also has the property
that the parity of the two-qubit state in $\mathcal{C}_2$ can be
directly measured using the coupler energy splitting, which could be
useful for error-correction.
\bibitem{appcheck} We simulated the full single-qubit dynamics
($\pm$16 charge states) under the action of flux and microwave
pulses; pulse parameters were chosen to keep nonadiabatic and
off-resonant microwave excitation out of the lowest two states below
$10^{-5}$. We checked the $n_{ph}=1$ truncation of resonator photon
states by simulating up to $n_{ph}=3$.
\bibitem{numvals} We assumed the following physical values:
Data qubits: $E_\textrm{J}/h=175$ GHz, $I_\textrm{p}=250$ nA,
$E_\textrm{J}/E_\textrm{C}=40, \beta=0.925$, and
$\Phi_\Delta=371\;\textrm{m}\Phi_0$; Coupler qubit:
$E_\textrm{J}/h=350$ GHz, $I_\textrm{p}=500$ nA,
$E_\textrm{J}/E_\textrm{C}=160,\beta=0.375$;
($\Phi_\epsilon,\Phi_\Delta)=(0,179)$ and
$(5.44,0)\;\textrm{m}\Phi_0$ for the coupled and uncoupled states.
Then, $M_\textrm{iC}=$ 10 pH gives $J_\textrm{C}/h\approx 0.3$ GHz.
\bibitem{frequencies} For each $\epsilon$ and $\Delta$ we used 40 independent
fluctuators, with average switching rates spaced logarithmically
from $\nu_\textrm{ir}=1$ Hz to $\nu_\textrm{uv}=0.2$ GHz, and
relative amplitudes chosen to maximize convergence to a 1/f
spectrum. Net amplitudes are determined by the slopes
$d\epsilon/d\Phi_\epsilon$ and $d\Delta_i/d\Phi_\Delta$. Error rates
do depend on $\nu_\textrm{ir}$ as expected~\cite{ithier05}, but are
independent of $\nu_\textrm{uv}$ for $\nu_\textrm{uv}\ge 0.2$ GHz.
\bibitem{coop} We explicitly neglect any cooperative effects in
decay.
\bibitem{resolution} Our simulations also allow us to determine the control-pulse
requirements for $\mathcal{E}\sim 10^{-5}$. For $\Omega_\upmu (t)$,
$\epsilon_\textrm{C}(t)$, and $\Delta_\textrm{C}(t)$: 8 bit vertical
accuracy, 1 ps edge timing; for $\Delta_1(t)$ and $\Delta_2(t)$, 14
bit, 0.2 ps. The latter pair can be relaxed by reducing the
qubit-resonator detuning; however, this would also increase
off-resonant mixing between them and reduce the resonator lifetime.
Removing the resonators entirely is also possible, and will be
considered in a future publication.
\bibitem{nonad} The error ``floor" for $A_\omega\rightarrow 0$ is caused by residual
nonadiabaticity in the qubit-resonator state transfer.
\end{thebibliography}
\end{document}